# Experimental observation of chaos-chaos intermittency types in spherical Couette flow

D. Zhilenko,[1,a)]  O. Krivonosova[1]

[1]*Institute of mechanics of Moscow State University, Moscow, 119192, Russia*

Flows between concentric, counter rotating spherical boundaries are under investigation in the gap with size equal to inner sphere radius. Outer sphere rotational rate is fixed, while inner sphere rotational rate has time periodic variations. The amplitudes and frequencies of these variations are small relative to both spheres averaged rotational rates. With amplitude increase transition from initial periodical flow to chaos is occurred. To determine state of the flow time series of azimuthal velocity, taken with laser Doppler anemometry, were used. We demonstrate appearance of flow states in the form of chaos-chaos and cycle-chaos-chaos intermittency. A procedure is considered which allow quantitatively confirm distinct properties of different patterns of time alternating flow state with intermittency.

Transitional processes to turbulence often are accompanied by intermittency in the form of laminar and turbulent bands, alternating in space and/or time. Intermittency was observed in transitional flows in tubes [1] and in flows, induced by the motion of the walls in opposite directions. Well-known cases of such flows are the plane Couette flow (PCF) – between flat boundaries [2,3], circular (CCF) – between rotating coaxial cylinders [4,5] and spherical [6,7] (SCF) – between coaxial rotating spheres. The possibility of coexistence in the spatial/temporal structure of distinct bands with

---

[a)] Author to whom correspondence should be addressed. Dr. Dmitry Zhilenko, E=mail: jilenko@imec.msu.ru

different turbulence intensity was shown for PCF and CCF. In SCF only temporal alternating of laminar and chaotic states was found – each state occupies the whole gap [6, 7].

Flow states with the intermittency may be divided by turbulent intensity distribution in space and/or time. Thus, in numerical investigation of PCF [3] was observed coexistence of different kinds of turbulent patterns, titled as uniform turbulence, intermittent turbulent-laminar patterns, localized states and laminar flow, distinguished from each other generally by spatial wave numbers and probability distribution function structure. In contrast to numerical results [3], available up to now experimental data demonstrate most likely modulated with constant frequency turbulence instead of different kinds of turbulent flow patterns. In CCF experiments [5] from time series of measured local velocity non-uniform in space and time turbulent flows were detected with periodical variation of turbulent intensity. Amplitude values of oscillations are comparable with the intensity of turbulent fluctuations. Results of SCF experiments [8] demonstrate, using torque time series, existence of turbulent fluctuations with different features, imposed on periodical oscillation. Amplitude values of oscillations were considerably larger than intensity of turbulent fluctuations.

In this letter we study the possibility of obtaining temporally alternating turbulence in SCF with modulated in time boundary conditions. Transitions to chaos in CCF, induced by sharp changing of one of the boundaries rotational rate, are well known beginning from [9].

Later similar procedure was used for transitions to chaos investigations in the flow around torsionally oscillating sphere [10]. For our experiments in SCF we use the same technique – transition to turbulence occurs with the growth of periodical modulation amplitude of inner sphere rotational rate, averaged angular velocities of both spheres remain constant. Steady SCF with stationary boundary conditions is determined by three dimensionless parameters. They are the Reynolds numbers for inner ($Re_1 = \Omega_1 r_1^2 / \nu$) and outer ($Re_2 = \Omega_2 r_2^2 / \nu$) spheres and relative gap size $\delta = (r_2 - r_1)/ r_1$. Here $r_1$, $r_2$ are the radii and $\Omega_1$, $\Omega_2$ the angular velocities of inner (index 1) and outer (index 2) spheres, $\nu$ is the kinematic viscosity of the fluid in the gap. The modulation of inner sphere angular velocity takes place in accordance with the law: $\Omega_1(t) = \Omega_{10}(1+A \sin (2\pi f t + \Phi_1))$, where A and f are the amplitude and frequency of modulation, $\Omega_{10}$ – averaged magnitude of inner sphere angular velocity, initial value of phase $\Phi_1$ is undefined. In case of modulation we use modified Reynolds number $Re_m = (A\, Re_1)(\delta_1/r_1)$, where $\delta_1 = (2\nu/2\pi f)^{1/2}$. Amplitudes and frequencies of modulation are small relative to their average values: $A \leq 0.2$, $2\pi f/\Omega_{10} \leq 0.1$.

Experimental setup represents two independently rotating coaxial plexiglas transparent spheres with $r_1 = 75$mm and $r_2 = 150$mm, $\delta = 1$. Gap between spheres is filled by silicon oil with high viscosity (near $50 \cdot 10^{-6} m^2/s$ at $22^0 C$). Small amount of aluminium flakes is added in working fluid for flow visualization. Local azhimuthal velocity $u_\varphi(t)$ is measured near the outer sphere by laser Doppler anemometry with allowable velocity range $0.003 - 1$ m/$s$. Synchronous acquisition of the time series of flow velocity and inner sphere rotational rate continues more than $3582s$, that is $1140 - 1170$ outer sphere revolutions. Both spheres are placed into thermostat, filled with silicon oil to keep constant temperature in the layer with accuracy not less than $\pm 0.05^0 C$. As initial flow state we choose periodical flow, called in [11] "localized vortices". This flow state is formed in wide layer with counter rotating spheres [11], in this study we use the following combination of control parameters $Re_1 = 412.5 \pm 0.5$ and $Re_2 = -900 \pm 1$. The action of modulation at constant f magnitude begins from zero value of A, we increase A step by step (dA ≤ 0.01) up to transition to chaotic flow state, transition is occurred throw breakdown of initial flow state.

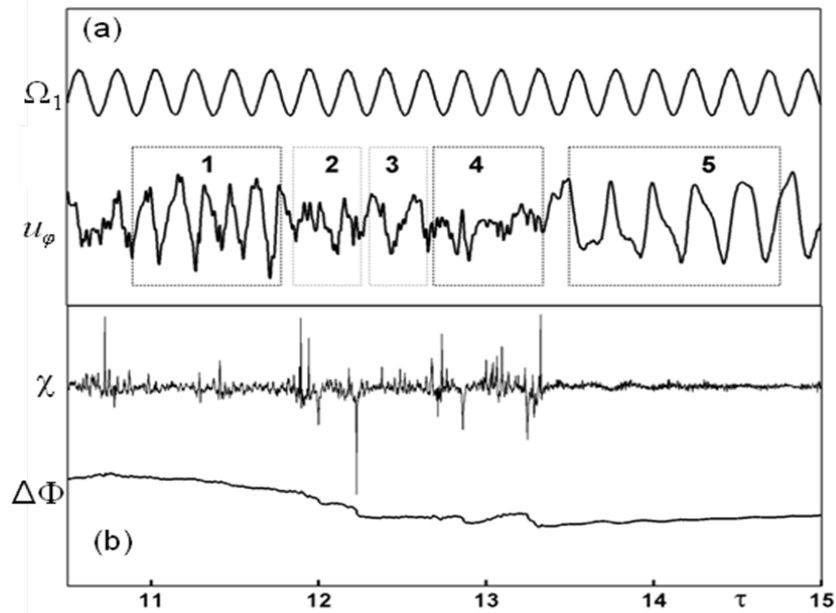

Fig. 1. (a) – smoothed signal of inner sphere angular velocity $\Omega_1$ (top) and raw signal of azimuthal velocity $u_\varphi$ (bottom) at $Re_m = 7.4$ depending on non-dimensional time τ common for both (a) and (b). (b) – raw instant frequencies difference χ (top) and smoothed (bottom) phase difference ΔΦ. All variables except τ are in arbitrary units.

In fig. 1(a) small portion of turbulent flow velocity time series is shown depending on non dimensional time $\tau = t\,(\nu / r_1^2)$. Different patterns of the signal with duration more than modulation period are clearly seen. Parts 1 and 3 represent distorted sinusoidal signal with frequency close to inner sphere modulation frequency. Parts 2 and 4 represent chaotic signal. Part 5, tentatively restricted from the right, represents modulated sinusoidal signal. Its frequency is less than modulation one and coincides with initial flow state frequency. Parts 1, 3 may be named as weak turbulence, parts 2, 4

– as strong turbulence, and part 5 – as laminar flow state. So, in this time series we observe cycle-chaos-chaos type intermittency. A question arises how one may distinguish patterns with different turbulence and what difference one may find in their properties. For this purpose we have chosen from a variety of known methods one, described in [12] and based on the introduction of phase not only for periodical signal, but for arbitrary one. According [12] the instant values of the phase $\Psi(t)$ of the signal $x(t)$ may be determined as $\Psi(t) = arctg\left(y(t)/x(t)\right)$, where $y(t)$ is orthogonal complement to $x(t)$ and is calculated as its Hilbert transform. Following this method we obtain instant values of phases, both for azimuthal flow velocity in separate point $\Phi_0(t)$ and for rotational rate of inner sphere $\Phi_1(t)$.

Results of application of this approach: phase difference $\Delta\Phi(t) = \Phi_1(t) - \Phi_0(t)$ and its derivative $\chi(t) = \partial(\Delta\Phi(t))/\partial t$ – instant frequencies difference – are present in fig.1(b). $\chi$ locates at the top, and $\Delta\Phi$, smoothed by eliminating $\pi$ jumps – at the bottom. $\Delta\Phi$ behavior (fig.1(b)) is slightly different for strong and weak turbulence. $\Delta\Phi$ for weak turbulence is smoother and keeps the slope sign: slope is near zero in part 3 and negative in part 1. Slope of $\Delta\Phi$ for strong turbulence does not keep sign. $\chi$ fluctuations for weak turbulence (parts 1, 3) are less than for strong ones (parts 2, 4). This means that in the case of weak turbulence synchronization exists between inner sphere and flow oscillations, what confirms by instant frequency adjustment. Laminar flow (part 5) differs from both turbulent ones by the least $\chi$ fluctuations level and by the constant positive slope of $\Delta\Phi$.

Using these differences, one can separate fragments with different flow states from smoothed $\Delta\Phi$ data, and define quantitative difference between them using $\chi$ data. The difference in root mean squares (rms) of $\chi$ for different patterns is demonstrated in

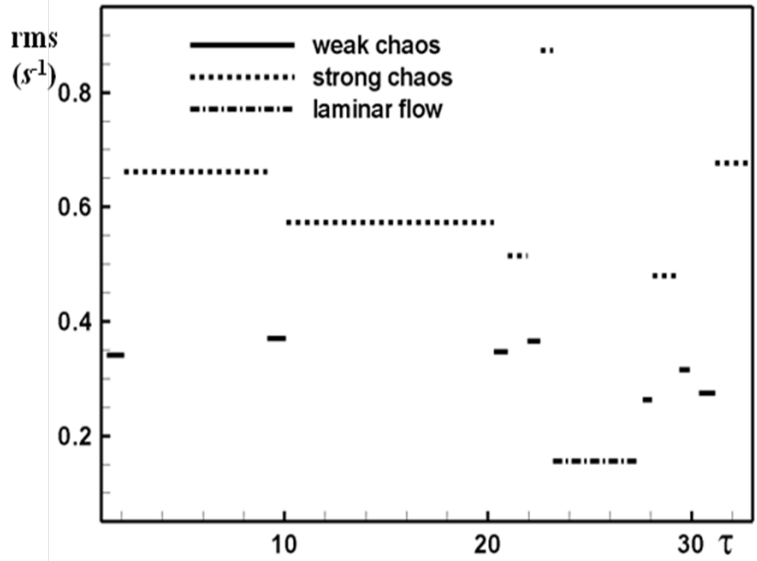

Fig.2. $\chi$ rms for different parts of the flow at Re $_m$ = 7.88.

Fig. 2. Rms level larger for all fragments with strong turbulence, one and a half times less for weak turbulence and minimal for laminar flow state. So we have quantitative difference for discriminating flow states.

In our experiments we have received three time series with cycle-chaos-chaos intermittency and then, by increasing Re$_m$, three time series with chaos-chaos intermittency. Relative time length of laminar parts $\lambda_{la}$ decreases with the growth of Re$_m$ in the first case, and disappear for the chaos-chaos intermittency (fig. 3). Relative time lengths of strong $\lambda_s$ and weak $\lambda_w$ turbulence in first case (three minor values of Re$_m$) demonstrate existence of local extremes depending on Re$_m$. In second case (another three values of Re$_m$) this dependence is linear. In the plot of $\chi$ rms, calculated along the whole time series, there is a local minimum in the point of transition from one case to another. Thus both kinds of intermittency differ from each other.

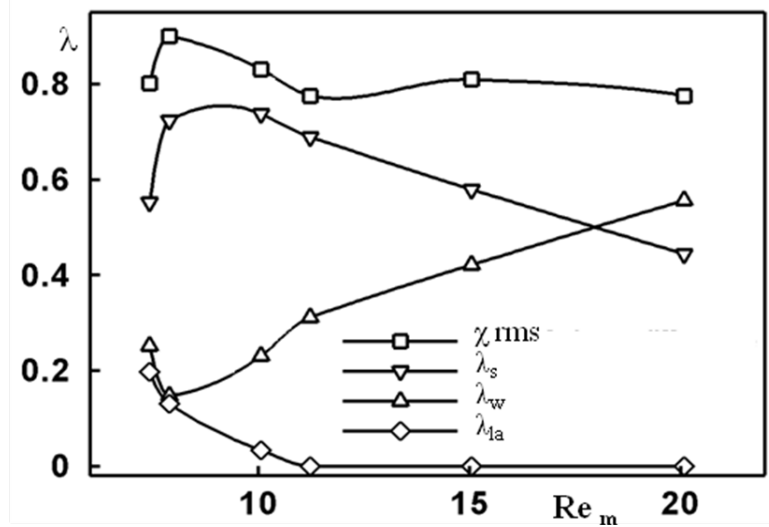

Fig.3. χ rms (arbitrary units) and relative time lengths of different kinds of flow states: strong turbulence – $\lambda_s$, weak turbulence – $\lambda_w$, laminar – $\lambda_{la}$ in time series depending on $Re_m$

In summary, presented in this letter results show the possibility of formation of complex turbulent flows with arbitrary alternating in time behavior. The reason of such behavior is the action of inner sphere rotation rate modulation, which may be considered as external force relative to the flow in the gap. This external periodical force may synchronize the flow, what means frequency and/or phase lock in. It is likely, that weak and strong turbulent fragments in time series of measured velocity are distinguished from each other by synchronization level, and different magnitudes of χ rms confirm that. On the other hand, the laminar flow states presence results in increasing phase difference as it may be concluded from $\Delta\Phi$ curve inclination in the part 5 in fig. 1. So, $Re_m$ growth promotes to forced synchronization, and laminar patterns existence resists it. Competition of these two processes gives complex dependence of weak and strong turbulence time duration on $Re_m$. When laminar patterns disappear, only one process influences on time durations $\lambda_s$ and $\lambda_w$, and that may explain their linear behavior during chaos-chaos intermittency, fig. 3. Further investigations are need for more detailed explanation of the modulation influence on intermittency formation.